\def\funp{{I\!\!P}}
\def\xp{x_{{I\!\!P}}}
\newcommand{\be}{\begin{equation}}
\newcommand{\ee}{\end{equation}}
\newcommand{\beeq}{\begin{eqnarray}}
\newcommand{\eeeq}{\end{eqnarray}}
\documentstyle[12pt,epsfig]{article}
\newlength{\dinwidth}
\newlength{\dinmargin}
\begin{document}

\begin{flushright}
INP Cracow  1754/PH\\
DTP/97/64\\
July 1997
\end{flushright}

\vskip 4cm
\begin{center}
{\large \bf Reggeon and pion contributions}
\vskip 0.1cm
{\large \bf in semi-exclusive  diffractive processes at HERA}
\vskip 1.5cm
{
K. Golec-Biernat$^{a,b}$ 
J. Kwieci\'nski$^{a}$ 
and A. Szczurek$^{a}$ 
}
\vskip0.5cm
{\it $^{a}$ H. Niewodnicza\'nski Institute of Nuclear Physics,\\
ul.Radzikowskiego 152,
PL-31-342 Krak\'ow, Poland}\\
\vskip0.2cm
{\it
$^{b}$Department of Physics, University of Durham, Durham DH1 3LE, UK}
\end{center}
\vskip3cm

\begin{abstract}
A detailed analysis of semi-exclusive diffractive processes
in $ep$ DIS at HERA, with the diffractive final states in
the forward direction is presented. The contributions of
the subleading $f_2$, $\omega$, $a_2$, $\rho$ reggeons and
the pion exchanges to the diffractive structure function with
the forward proton or neutron are estimated.
It is found that the $(a_2, \rho)$ reggeons
are entirely responsible for the forward neutron production at
$\xp < 10^{-3}$. The $\pi N$ production in the forward region is
estimated using the Deck mechanism. The significance of this reaction
for the processes measured at HERA, especially with the leading neutron,
is discussed.

\end{abstract}

\setcounter{page} {0}
\thispagestyle{empty}

\newpage
The diffractive processes in deep inelastic scattering observed
recently at the $ep$ collider HERA at DESY by the H1 and ZEUS
collaborations \cite{H1,ZEUS}, were interpreted in terms of
the exchange of the leading Regge trajectory corresponding to
the ''soft'' pomeron \cite{CKMT95,GK95,GS96,KKK95} (for
alternative explanations see \cite{NZ92,BH95}).
In this approach, first proposed
by  Ingelman and Schlein\cite{IS85}, the diffractive interaction
is treated as a two-step process: an emission of the pomeron from a
proton and subsequent hard scattering of a virtual photon on partons
in the pomeron.
The idea that the pomeron has partonic structure was experimentally
supported by the hadron collider experiments \cite{UA8}.
In $ep$ scattering this idea is expressed through the factorization
of the diffractive structure function
\begin{equation}
\frac{dF_2^{D}}{dx_{\funp} dt}(x,Q^2,x_{\funp},t) =
f^{\funp}(x_{\funp},t) \, F_2^{\funp}(\beta,Q^2) \; ,
\label{diff4_P}
\end{equation}
where $f^{\funp}(x_{\funp},t)$ is the pomeron flux in the proton and
$F_2^{\funp}(\beta,Q^2)$ its DIS structure function. The kinematical
variables are defined as follows
\begin{eqnarray}
Q^2=-q^2~~,~~t=(p-p^{\prime})^2~~,~~x={Q^2\over 2pq}
~~,~~\beta={Q^2\over 2q(p-p^{\prime})}
~~,~~x_{I\!\!P}={x\over \beta}~.
\label{invar}
\end{eqnarray}
and  $q=p_e-p_e^{\prime}$, where  $p_e,p_e^{\prime},p$  and
$p^{\prime}$ are the momenta of the initial and final electron,
initial and recoiled proton respectively.

The pomeron structure function $F_2^{\funp}(\beta,Q^2)$ is
related to the parton distributions in the pomeron in a full analogy
to the nucleon case. Because the pomeron carries
the vacuum quantum numbers,
the number of independent distributions is smaller than for
the proton. Recently the partonic structure was
estimated \cite{CKMT95,GK95,GS96}, and independently fitted
to the diffractive HERA data \cite{H1,ZEUS}.
Contrary to the nucleon case a large gluon component of the pomeron
was found at $\beta \rightarrow 1$, \cite{H1paris}.

\section{Reggeons and pions in inclusive hard diffraction}

The new data shown by the H1 collaboration at
Warsaw conference \cite{Phillips96}
indicate breaking of the factorization (\ref{diff4_P}).
In order to describe this effect it was recently proposed
to include the contributions of subleading reggeons
\cite{Phillips96,GK96,NIK2}. In this
generalized approach the diffractive structure function can be written
as:
\begin{equation}
\frac{dF_2^{D}}{dx_{\funp} dt}(x,Q^2,x_{\funp},t) =
f^{\funp}(x_{\funp},t) \, F_2^{\funp}(\beta,Q^2) \, + \,
\sum_{R} f^{R}(x_{\funp},t) \, F_2^R(\beta,Q^2) \; ,
\label{diff4_PR}
\end{equation}
where $f^{R}$ and $F_2^R$ are reggeon flux and structure function
respectively.
\footnote{To be precise the term "diffractive processes" applies
only to processes described by the pomeron exchange.
For simplicity we shall use the same terminology for the non-pomeron
reggeon exchanges including processes with the forward neutron in
the final state which correspond to $I=1$ exchange.}

In the present paper we take the pomeron flux factor of the form
given in \cite{GK95,CHPWW95}
\begin{equation}
f^{\funp}(x_{\funp},t) =
{N \over 16 \pi} \; x_{\funp}^{1-2 \alpha_{\funp}(t)} \; B_{\funp}^2(t)
\; ,
\label{pomeron_flux}
\end{equation}
where $\alpha_{\funp}(t) = 1.08 + (0.25~GeV^{-2}) t$ is the "soft"
pomeron trajectory,  $B_{\funp}(t)$ describes the pomeron-proton
coupling and $N = 2/\pi$, following the discussion in \cite{CHPWW95}.
In analogy to the pomeron the subleading reggeon flux factors
are parametrized as
\begin{equation}
f^R(x_{\funp},t) =
  {N \over 16 \pi} \;
 x_{\funp}^{1-2\alpha_R(t)} \;
 B_{R}^2(t) \; |\eta_R(t)|^2 \;,
\label{reggeon_flux}
\end{equation}
where $\alpha_R(t)=\alpha_R(0)+\alpha_R^{'}~t$ is the reggeon
trajectory,
$B_{R}(t)$ describes the coupling of the reggeon to the proton and
is assumed to have the form
 $B_{R}(t) = B_{R}(0) \exp \left({t \over 2 \Lambda_R^2} \right)$ with
$\Lambda_R=0.65~GeV$,
as known from the reggeon phenomenology in hadronic reactions
\cite{Regge}.
The function $\eta_R(t)$ is  a signature factor \cite{Collins};
$|\eta_R(t)|^2 = 4\,\cos^2(\pi \alpha_R(t)/2)$
and
$|\eta_R(t)|^2 = 4\,\sin^2(\pi \alpha_R(t)/2)$
for even and odd signature reggeons respectively.

The analysis done in \cite{GK96} for the isoscalar reggeons
$(f_2,\omega)$
shows that this contribution to the diffractive structure function
becomes important for $x_\funp > 0.01$.
In the present paper we extend this analysis by including the
isovector subleading reggeons $(a_2,\rho)$.
Although the isovector reggeons contribution to the total diffractive
structure function is small, it becomes important for the
processes with leading neutron in the final state.
We will show that for small values of $x_{\funp}$ this is the dominant
contribution.
In addition, we add to our analysis the one-pion exchange contribution
which is expected to be important at large $x_\funp$ values.
The pionic contribution to deep inelastic lepton-proton scattering
(Sullivan process) was found to provide a parameter-free description of
the experimentally observed $\bar u - \bar d$ asymmetry in the proton
\cite{HSS96}.

In the past several phenomenological methods were used to determine
the contribution of subleading reggeons in hadronic reactions.
Inspired by the success of a recent parametrization of the total
hadronic cross sections (see for instance \cite{LL93})
by Donnachie and Landshoff \cite{DL92} we parametrize total cross
sections as:
\begin{eqnarray}
\nonumber
\sigma_{tot}^{\bar p p}(s) &=& \sigma_{P}(s) +
\sigma_{f_2}(s) + \sigma_{\omega}(s) +
\sigma_{a_2}(s) + \sigma_{\rho}(s) \; , \\
\nonumber
\sigma_{tot}^{p p}(s) &=& \sigma_{P}(s) +
\sigma_{f_2}(s) - \sigma_{\omega}(s) +
\sigma_{a_2}(s) - \sigma_{\rho}(s) \; , \\
\nonumber
\sigma_{tot}^{\bar p n}(s) &=& \sigma_{P}(s) +
\sigma_{f_2}(s) + \sigma_{\omega}(s) -
\sigma_{a_2}(s) - \sigma_{\rho}(s) \; , \\
\sigma_{tot}^{p n}(s) &=& \sigma_{P}(s) +
\sigma_{f_2}(s) - \sigma_{\omega}(s) -
\sigma_{a_2}(s) + \sigma_{\rho}(s) \; .
\end{eqnarray}
Assuming the Regge-like energy dependence of each contribution
\begin{equation}
\sigma_R(s) = \sigma_R(s_0) \, \left( \frac{s}{s_0} \right)^{1-\alpha_R(0)}
\end{equation}
and the same intercept for each reggeon, one finds:
$\alpha_R(0)$ = 0.5475, $B_{f_2}^2(0)$ = 75.49 mb, $B_{\omega}^2(0)$ =
20.06 mb, $B_{a_2}^2(0)$ = 1.75 mb and $B_{\rho}^2(0)$ = 1.09 mb. One
clearly sees the following ordering
\begin{equation}
B_{f_2}^2(0) > B_{\omega}^2(0) \gg B_{a_2}^2(0) \sim B_{\rho}^2(0) \; ,
\label{order}
\end{equation}
which implies dominance of isoscalar reggeons over isovector ones.
One should remember, however, that the contributions of the latter will
be enhanced by the appropriate isospin Clebsch-Gordan factor of 3, when
going to  deep-inelastic $ep$ scattering and including both forward
proton and neutron in the final state.

In Fig.1 we show the $f_2$, $\omega$, $a_2^{0}$ and $\rho^{0}$ reggeon
flux factors as a function of $x_\funp$
integrated over $t$ up to the kinematical limit
$t_{max}(x_\funp)=-\frac{m^2_Nx_\funp^2}{1-x_\funp}$.
The flux factors reflect the ordering of the normalization constants
(\ref{order}) and a small difference in shapes at large values of
$x_\funp$ is caused by the fact that the signature factors are
different for the positive and negative signature reggeons.
A more complicated structure at larger $x_\funp> 0.1$
(not shown in Fig.1) comes from the integration limits
$t_{max}(x_\funp)$. One should remember, however, that the Regge
(high-energy) approximation applies exclusively to small
$x_\funp$ values, definitively smaller than $0.1$.
For comparison we also present in Fig.1 the pomeron (solid line) and
pion (dashed line) flux factors.

The pion flux factor $f(x_\pi,t)$ has the following form resulting
from the one pion exchange model
\begin{equation}
f_{\pi N} (x_\pi,t) = \frac{g_{p \pi^0 p}^2}{16\pi^2}
x_\pi \frac{ (-t) |F_{\pi N}(x_\pi,t)|^2 } {(t - m_{\pi}^2)^2} \; ,
\label{splitpiN}
\end{equation}
where $g_{p \pi^0 p}$ is the pion-nucleon coupling constant and
$F_{\pi N}(x_\pi,t)$ is the vertex form factor which accounts for
extended nature of hadrons involved. The variable $x_\pi$ is
a fraction of the proton longitudinal momentum carried by the
pion, equal to $x_\funp$ defined in (\ref{invar}).
The form factors used in meson exchange models are usually taken to
be a function of $t$ only. The form (\ref{splitpiN}) corresponds to
setting the pion Regge trajectory $\alpha_{\pi}(t)=0$.
Here for simplicity we shall use the same exponential form factor as
for the reggeon-nucleon coupling in formula (\ref{reggeon_flux}).
The cut-off parameter $\Lambda_\pi= 0.65~GeV$ gives the same
probability of the $\pi N$ component as that found in \cite{HSS96}.
As seen in Fig. 1 the pionic contribution becomes important
only at $x_{\funp} > 0.05$.
It was shown recently, however, that this component gives rather small
contribution to large rapidity gap events \cite{PSI96}.

The pomeron and reggeon structure functions, $F_2^{\funp}$ and $F_2^{R}$
in relation (\ref{diff4_PR}),
are related to the parton distributions in the pomeron
and reggeons in a conventional way, and can be estimated
using the "soft" pomeron interaction properties and "triple-Regge"
phenomenology \cite{CKMT95,GK95,GS96,GK96}.
They can also be constrained by the fit to diffractive DIS data from
HERA \cite{H1paris,Phillips96}.
In the method presented in \cite{GK95,GK96} the small $\beta$
behaviour is the same for both the pomeron and reggeon structure
functions:
\begin{equation}
F_2^R(\beta) = A_R \; \beta^{-0.08}  \; .
\label{pbeta}
\end{equation}
The coefficients $A_\funp$ for the pomeron
and $A_R$ for the reggeon are related to the "triple-Regge"
$\funp \funp \funp$ and $R R \funp$ couplings
respectively, and their ratio
%
\begin{equation}
\label{cenh}
C_{enh} = \frac{A_R}{A_\funp}~,
\end{equation}
is varied in the interval  $1< C_{enh} < 10$,
as suggested by the "triple-Regge" analysis of
inclusive processes in hadronic reactions \cite{FFOX}. We extrapolate
parametrization (\ref{pbeta}) to the region of large $\beta$
multiplying the r.h.s. of Eq. (\ref{pbeta}) by $(1-\beta)$.

In contrast to the reggeon case the structure function of
the pion at large $\beta$ is fairly well known from
the analysis of the pion-nucleus Drell-Yan data.
The region of small $\beta$, however, cannot be constrained
by the available experimental data.
In our calculations we shall take the pion structure function
as parametrized in \cite{GRV92}, which in the region of $\beta > 0.1$
properly describes the pion-nucleus Drell-Yan data.

Having fixed all ingredients we calculate the contributions
of different reggeons to the diffractive structure function
integrated over $t$
\begin{eqnarray}
\label{integF}
\nonumber
F_2^{D(3)}(x,Q^2,x_\funp) &=& f_{\funp}(x_\funp)~F_2^{\funp}(\beta,Q^2)
\\ \nonumber
\\
&+& \sum_{R} f_{R}(x_\funp)~F_2^{R}(\beta,Q^2)
               ~+~ \sum_{\pi} f_{\pi N}(x_\funp)~F_2^{\pi}(\beta,Q^2)~.
\end{eqnarray}
%
In Fig.2 we show the function $x_{\funp} F_2^{D(3)}(x_{\funp})$
at $Q^2$ = 4 GeV$^{2}$,
for two extreme values of $\beta = 0.01$ and $0.7$ and
two values of the parameter $C_{enh} = 2$ and $10$,
calculated with the quark distribution functions in the pomeron from
\cite{GK95}.
The evident rise of the $F_2^{D(3)}$ structure function at
$x_{\funp} > 0.02$ is an effect of the subleading reggeons and pions;
the reggeon contribution being considerably bigger for $C_{enh}$ = 10,
whereas for $C_{enh}$ = 2 the pion contribution being equally
important (compare dashed and dash-dotted lines).
We expect that a future HERA data will allow to fix the presently
unknown parameter $C_{enh}$.
It should be noted, however, that the rise of
the structure function is not far
from the region where the Regge parametrization is not expected
to be valid, therefore some caution is required in the analysis of
experimental data.

\section{Fast forward neutron production}

The recent discovery of large rapidity gap events at HERA is based
on the inclusive analysis of rapidity spectra of particles
which entered the main calorimeter. It is expected that these
events are associated with the production of a fast baryon, the proton
being probably the dominant case.
The installation of the leading proton spectrometer and forward
neutron calorimeter opens up a possibility to analyze the diffractive
events more exclusively.
In particular these experimental efforts will allow to answer the
question what fraction of diffractive deep inelastic events is
associated with the emission of fast forward proton and neutron.

The model presented above allows to separate the diffractive
structure function (\ref{integF}) into two distinct contributions
\begin{equation}
F_2^{D(3)}(\beta,x_{\funp},Q^2) =
\Delta^{(p)}
(\beta,x_{\funp},Q^2) +
\Delta^{(n)}
(\beta,x_{\funp},Q^2) \; ,
\label{pncontri}
\end{equation}
where the upper indices $p$ and $n$ denote the leading proton or neutron
observed in the final state respectively. In order to calculate the
functions $\Delta^{(p)}$ and $\Delta^{(n)}$ we make use of the isospin
symmetry for the flux factors of the corresponding reggeons:
\begin{equation}
f_{\rho^{+}}(x_\funp) = 2~f_{\rho^{0}}(x_\funp)
~~~~~~~~~~,~~~~~~~~~~~
f_{a_2^{+}}(x_\funp) = 2~f_{a_2^{0}}(x_\funp)~.
\end{equation}

In Fig 3. we show the proton and neutron contributions
to the diffractive structure function (\ref{pncontri}),
marked by the solid lines.
As expected the proton contibution dominates
over the neutron one almost in the whole range of $x_{\funp}$ because
of the dominace of the pomeron contribution in $\Delta^{(p)}$, absent
in $\Delta^{(n)}$.
The ratio $\Delta^{(n)}/\Delta^{(p)}$ becomes significant ($\sim 0.1$)
only for $x_{\funp} \approx 0.1$, where the pomeron exchange
is suppressed. In this case the reggeon and pion contributions
come into play. However, as we have stressed before, this region needs a
careful treatment since it might already be unsuitable for
the Regge analysis.

The lower curves in Fig.3 shows different contributions to the
fast forward neutron production. For large values of
$x_{\funp}$ the $\pi^{+}$  exchange process,
marked by the dash-dotted line,
is the dominant effect for the fast neutron production.
The $a_2$ and $\rho$ isovector reggeon exchange
contributions, shown by the dashed line, are by the order of magnitude
smaller.
The situation changes dramatically when $x_{\funp}$ is getting
smaller. The reggeon exchanges are almost entirely
responsible for the fast neutron production for $\xp < 10^{-3}$.

Is it possible to identify these contributions experimentally?
In Fig.4 we present the cross section for the fast neutron
production integrated over
$x$, $Q^2$ and $t$ for $C_{enh}$ = 10
\begin{equation}
\frac{d \sigma^{(n)}}{d x_{\funp}} =
\int_{0}^{x_{\funp}} dx  \int_{Q_{min}^{2}}^{Q_{max}^{2}(x)} dQ^2
\int_{t_{min}}^{t_{max}(x_R)} dt
\; \frac{d \sigma(ep \rightarrow enX)}{dx dQ^2 dx_{\funp} dt} \; .
\label{xR_distri}
\end{equation}
Here in order to be roughly consistent with the preliminary H1 data
\cite{FNC}, we have fixed $Q_{min}^2$ = 2.5 GeV$^2$ and
$t_{min}$ = -1 GeV$^2$. Although at $x_{\funp} \sim 0.1$ the charged
pion exchange dominates,
the $(a_2,\rho)$ reggeon exchanges take over at very small $\xp <$
10$^{-3}$.
This is exactly the region of $\xp$ where unexplained by the existing
Monte Carlo programs, enhanced strength has been observed \cite{FNC}.

\section{The Deck mechanism}

Up to now we have totaly neglected the contribution of diffractively
produced $\pi N$ and $\pi \pi N$ states. The latter one becomes
important only for larger values of $\xp$ \cite{HNSSZ96}. For
small $\xp$
relevant here only the $\pi N$ contribution is of interest.

Let us try to estimate the contribution of diffractively
produced $\pi N$ state in the forward region
(with respect to the proton beam).
In hadronic reactions the $\pi N$ component is known to be
produced dominantly by the Deck mechanism (see for instance
\cite{AG81}). The Deck mechanism can be generalized to the case
of lepton DIS. In analogy to the hadronic diffractive
production there are three lowest order diagrams leading to
the $\pi N$ final state (see fig.5.1 in \cite{AG81}).
In hadronic reactions for small masses of the pion-nucleon system
there is almost exact cancellation of the "s-" and "u-" channel
diagrams \cite{AG81}.\footnote{Here "s-", "u-" and "t-" channel
diagrams correspond to the diagrams with single particle exchanges in
the "s-", "u-" and "t-" channels of the "process"
$\funp p \rightarrow \pi N$.}
Both "s-" and "u-" channel diagrams are expected
to have much smaller slope in $t$ distribution. Therefore at small $t$,
relevant for the FNC measurements, the contribution of "u-" and
"s-" channel diagrams is expected to be rather small. The same argument
applies to the diffractive production of proton resonances.
The decay of protonic resonances into neutron (proton) channel causes
both broadening of the $t$-distribution and an effective shift towards
larger $\xp$ values \cite{HNSSZ96}, making their identification
by the present ZEUS and H1 detectors rather inefficient.

In the following we shall therefore limit to the calculation of
the contribution of the "t-" channel diagram in which
the pomeron couples to the virtual pion.
In the simplest approximation the corresponding structure function
with leading proton or neutron can be written as:
\begin{equation}
\frac{d F_2^D}{dx_{\funp} dt} =
f_{\pi N}(x_{\pi},t) \cdot F_{2}^{{\funp}/\pi}(\beta,Q^2) \; ,
\label{Deck}
\end{equation}
where the flux factor here is exactly the same as the one for
the Sullivan process (\ref{splitpiN}) (with $x_{\pi}=\xp$).
The effective structure function
$F_{2}^{{\funp}/\pi}(\beta,Q^2)$ can be estimated as:
\begin{equation}
F_{2}^{{\funp}/\pi}(\beta,Q^2) =
\int_{\beta}^{1} dx_{\funp/\pi} \int_{-\infty}^{t'_{max}} dt'
\; f_{\funp/\pi}(x_{\funp/\pi},t') F_{2}^{\funp}(\tilde{\beta},Q^2) \; ,
\end{equation}
where $\tilde{\beta} \equiv \beta/x_{\funp/\pi}$.
In order to calculate the $f_{\funp/\pi}$ flux factor we assume
empirically established universality of the pomeron coupling to
hadrons which involves extra factor $2/3$ (two quarks for the pion
versus three quarks for the nucleon).

Since the same pion flux factor occurs
in Eq.(\ref{integF}) and Eq.(\ref{Deck}), it is sufficient to compare
the pion structure function $F_2^{\pi}$ and the
effective structure function $F_2^{\funp/\pi}$ to assess the importance
of the Deck mechanism in the fast proton or neutron production.
This is done in Fig. 5 where both functions
are shown together with their ratio at $Q^2$ = 4 GeV$^2$.
It becomes clear that at large $\beta$
the Deck mechanism for production of $\pi N$ states can safely be
neglected.
The Deck contribution can also be neglected in the
spectrum of neutrons as is shown by the dotted line in Fig. 3.

The situation at larger $x_{\funp}$ values
is less clear if one imposes the large rapidity gap condition.
While the Sullivan process does not lead to the large rapidity gap
events \cite{PSI96}, it may be not true for the Deck mechanism.
Because of identical flux factors for the Sullivan and Deck mechanisms
one expects very similar $\xp$ distributions.

An interesting and unique feature of the Deck mechanism is that it
contributes to large rapidity gap events even at
$x_{L}=1 - x_{\funp}$ much smaller than unity.
The preliminary ZEUS data \cite{Cartiglia97} seem to confirm our
approximation. An approximately constant in $x_{L}$ ratio of
\beeq
\label{nratio}
{\frac{{\rm large~rapidity~gap ~events~with~leading~neutron/proton}}
{{\rm total~ number~ of~ events~ with~ leading~ neutron/proton}}}
\eeeq
in the region $0.65 < x_L < 0.85$.
has been observed \cite{Cartiglia97}, consistent with our simple
estimate.
A detailed comparison with the experimental spectra
will require the inclusion of all experimental cuts and efficiencies
and goes beyond of the scope of the present paper.

\section{Summary}

We made a detailed analysis of semi-exclusive diffractive
processes in $ep$ DIS at HERA, with the diffractive
final states in the forward (with respect to the incoming proton)
direction.
The contributions of the subleading $f_2$, $\omega$, $a_2$ and $\rho$
reggeons as well as the pion exchanges
to the diffractive structure function with the forward proton
or neutron were estimated. In addition the $\pi N$ production
in the forward direction was computed using the Deck mechanism.
Our main results are summarize as follows.
The isoscalar reggeon exchanges together with the pion
exchanges can describe the Regge factorization breaking in
the inclusive DIS diffractive data observed by the H1 collaboration.

Their relative strength can be determined by the data.
The ${\pi}^{+}$ exchange is the dominant mechanism
for the fast forward neutron production at large $\xp$ values, while for
$\xp < 10^{-3}$ the isovector $(a_2, \rho)$ exchanges play the dominant role.
The later effect seems to explain an excess of neutrons produced in the
region of small $\xp$ values.
Finally, the contribution of the $\pi N$ production in the forward region,
described by the Deck mechanism, is found to be rather small in
the case of outgoing neutron ($N=n$) and even smaller in the case
of outgoing proton ($N=p$).
However this effect can be important for the forward neutron
production in the region of large values of $\xp$ and/or
small $\beta$ with an extra requirement of large rapidity gap.
The approximately constant ratio (\ref{nratio}),
observed by the ZEUS collaboration can be explained by the Deck
mechanism of the $\pi N$ production.

\vskip 1cm
{\bf Acknowledgments}
\par
We are indebted to Nicolo Cartiglia,
Lev Lipatov and Jullian P. Phillips for interesting
discussions.  
K.G-B thanks the Royal Society for the Fellowship and Department of
Physics of the University of Durham for its warm hospitality.
This research has been supported in part by the
Polish State Committee for Scientific Research grant
no.2 P03B 231 08 and 2 P03B 184 10,
and by the
Stiftung f\"ur Deutsch-Polnische Zusammenarbeit, project no.1522/94/LN


\newpage

\begin{figure}[htb]
   \vspace*{-1cm}
    \centerline{
     \psfig{figure=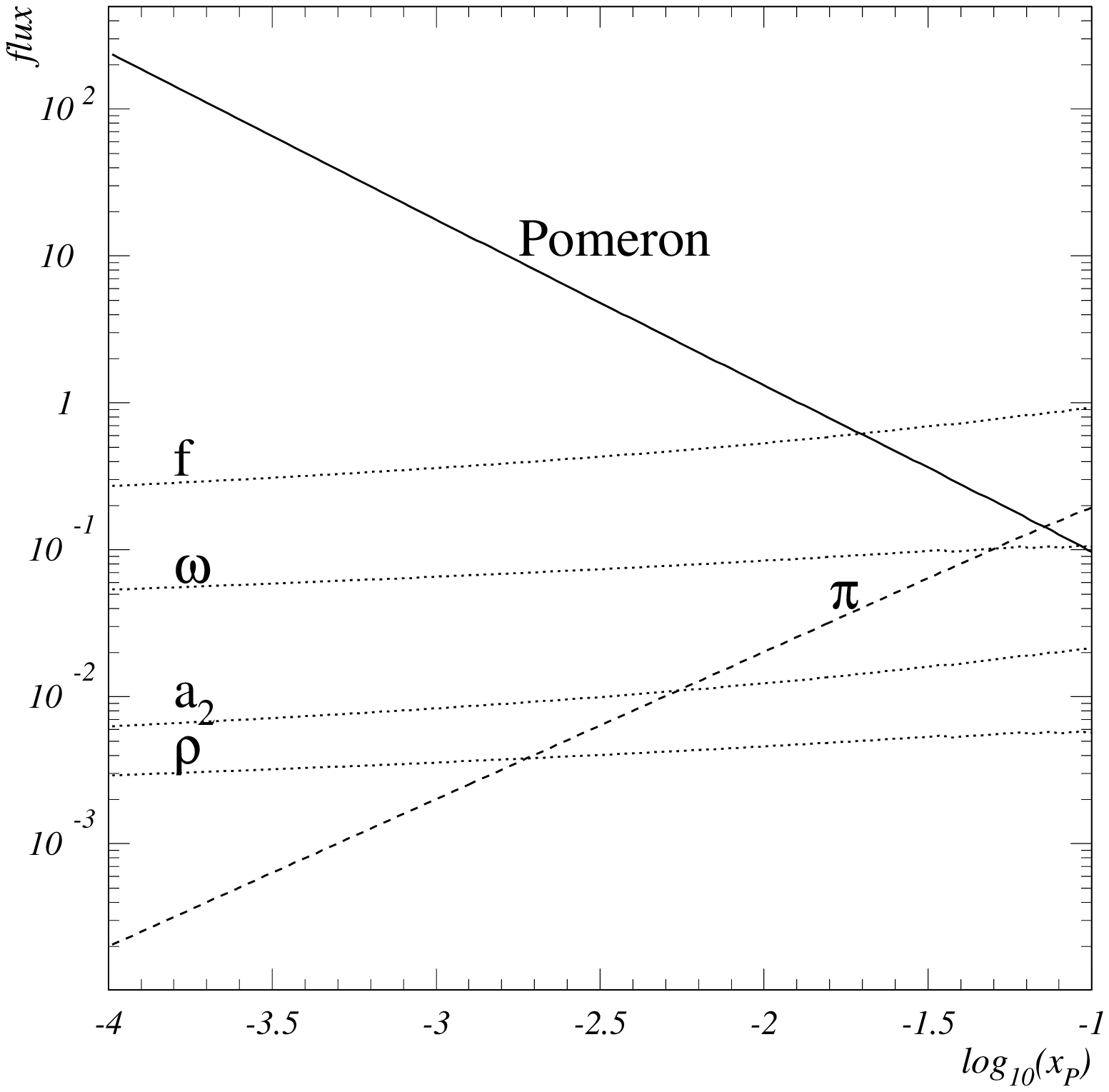,height=18cm,width=18cm}
               }
    \vspace*{-0.5cm}
     \caption{ 
The integrated over $t$ flux factors of $f_2$, $\omega$, $a_2$ and $\rho$
reggeons as a function of $x_\funp$. For comparison we present the pomeron
(dotted) and pion (solid) flux factors.
}
\end{figure}

\newpage

\begin{figure}[htb]
   \vspace*{-1cm}
    \centerline{
     \psfig{figure=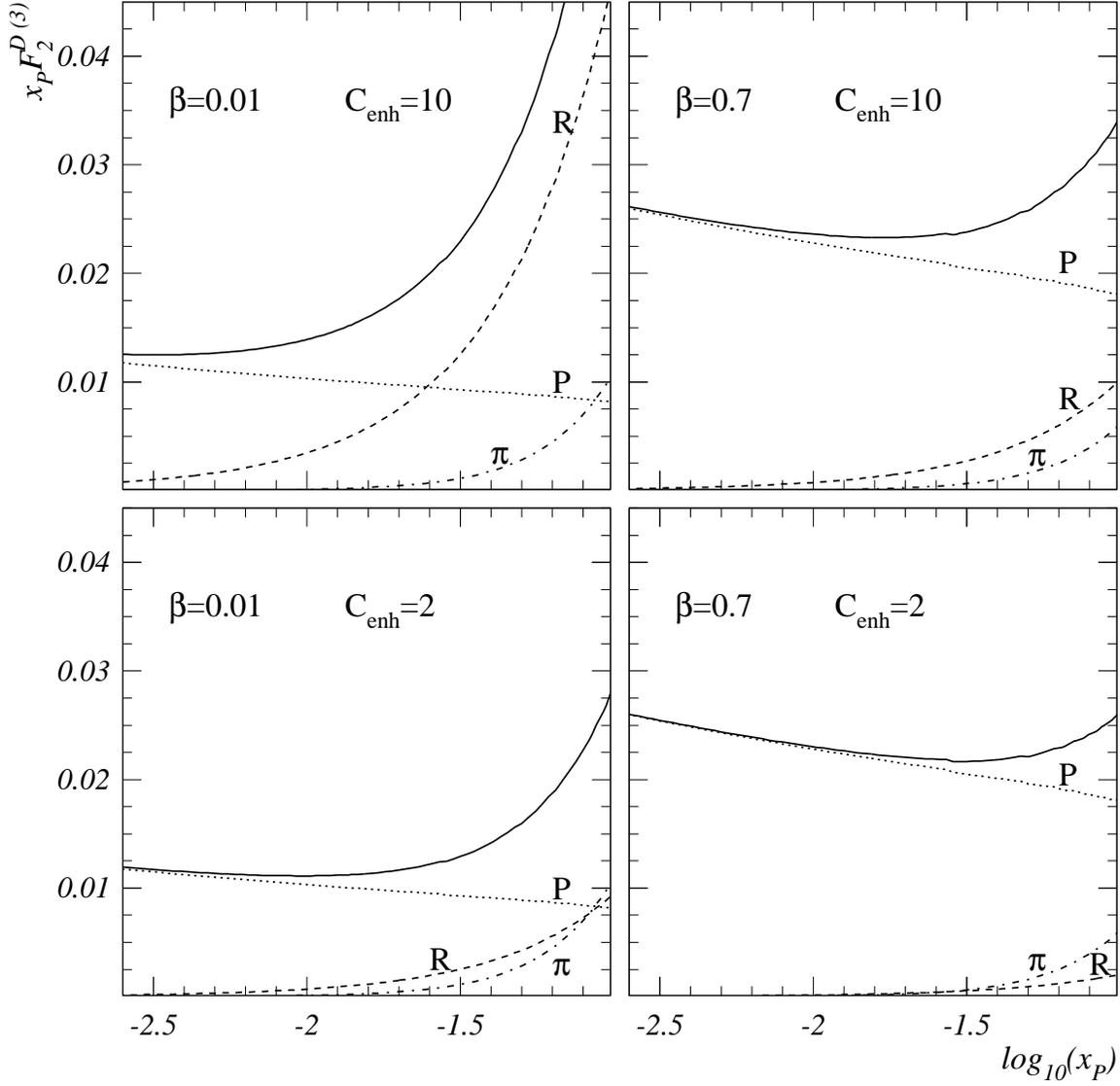,height=17cm,width=17cm}
               }
    \vspace*{-0.5cm}
     \caption{ 
The structure
function $x_{\funp}  \cdot F_2^{D(3)}(x_{\funp},\beta,Q^2)$ as a function 
of $\xp$ at $Q^2 = 4 GeV^2$ for two values $\beta$ and $C_{enh}$. 
The pomeron ($\funp$), reggeon (R) and pion ($\pi$) contributions 
to the total structure function (solid lines) are shown as seperate
curves.
}
\end{figure}

\newpage

\begin{figure}[htb]
   \vspace*{-1cm}
    \centerline{
     \psfig{figure=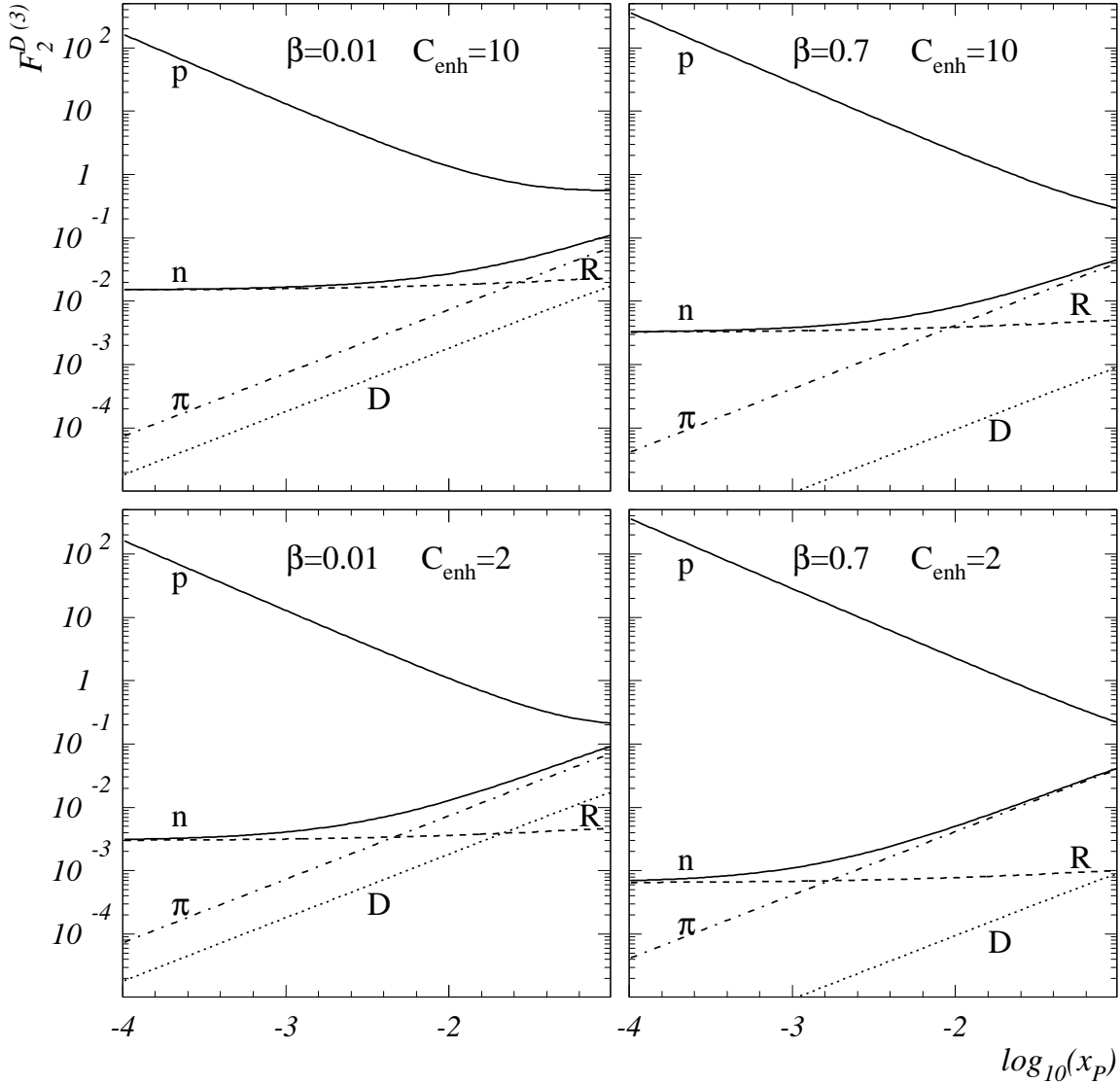,height=17cm,width=17cm}
               }
    \vspace*{-0.5cm}
     \caption{ 
The $\Delta^{(p)}$ and $\Delta^{(n)}$ contributions to the diffractive
structure function $F_2^{D(3)}$ as functions of $\xp$ for $Q^2 =4 GeV^2$
and for two values $\beta$ and $C_{enh}$ (solid lines). 
The contributions from
the $(a_2,\rho)$ (dashed lines) and  pion (dot-dashed lines) exchanges 
to $\Delta^{(n)}$ are shown. The $\pi N$ contribution for  neutron 
from the Deck model is marked by the dotted lines.
}
\end{figure}

\newpage

\begin{figure}[htb]
   \vspace*{-1cm}
    \centerline{
     \psfig{figure=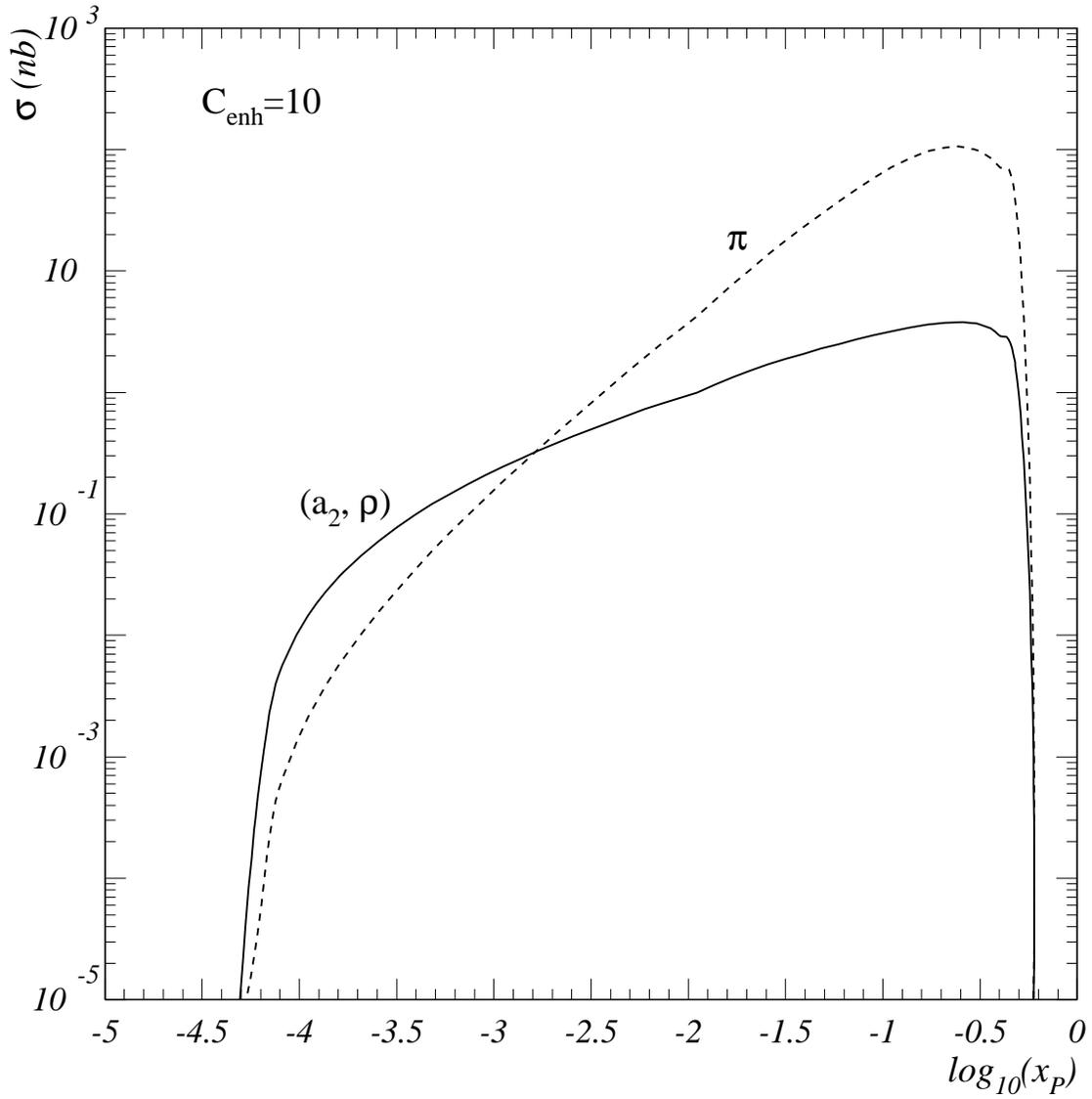,height=17cm,width=17cm}
               }
    \vspace*{-0.5cm}
     \caption{ 
Dominant contributions to the charge exchange
cross section ${d \sigma^{(n)}}/{dx_{\funp}}$ as a function
of $\xp$.
}
\end{figure}
\newpage

\begin{figure}[htb]
   \vspace*{-1cm}
    \centerline{
     \psfig{figure=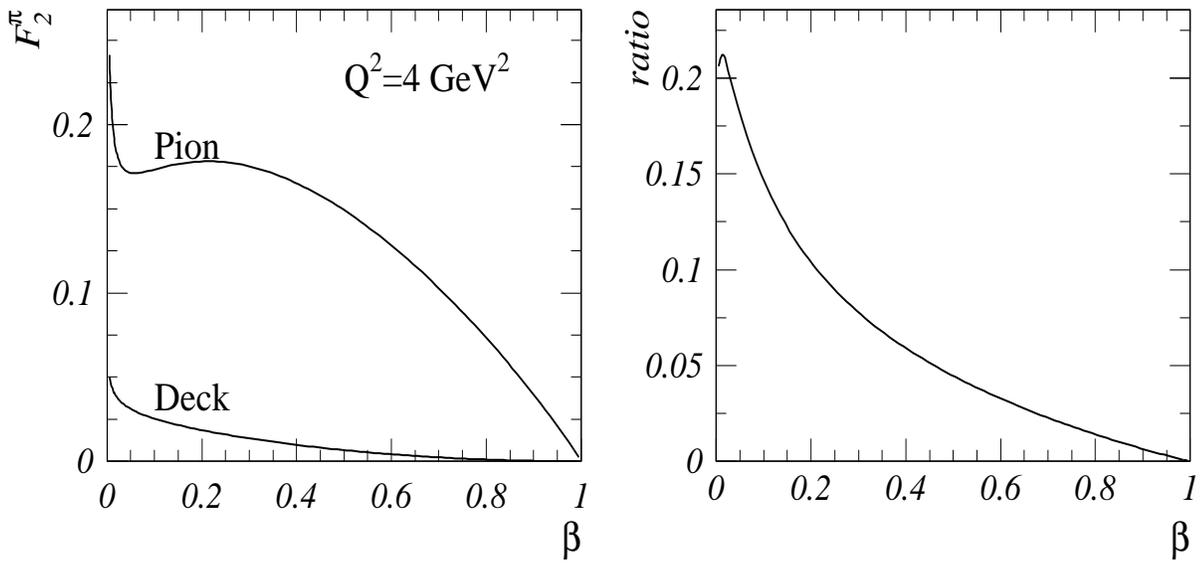,height=10cm,width=18cm}
               }
    \vspace*{-0.5cm}
     \caption{ 
The functions $F_2^{\pi}(\beta,Q^2)$ and $F_2^{{\funp}/\pi}(\beta,Q^2)$ versus
$\beta$ at $Q^2 = 4 GeV^2$ and their ratio.
}
\end{figure}

\end{document}